# Nanoscale characterization of the impact of beverages on the enamel surface of human teeth

Panpan Li, Chungik Oh, Hongjun Kim, Melodie Chen-Glasser, Gun Park, Albina Jetybayeva, Jiwon Yeom, Hoon Kim, Jeongjae Ryu and Seungbum Hong[*]

*Department of Materials Science and Engineering, Korea Advanced Institute of Science and Technology (KAIST), Daejeon, 34141, Republic of Korea*

---

[*] Corresponding author: seungbum@kaist.ac.kr.

**Abstract**

Here we quantitatively evaluate the early stages of mechanical and morphological changes of polished human enamel surfaces induced by soft drinks using atomic force microscopy. With an increase of the immersion time in soft drinks, we found a significant increase of surface roughness ($R_a$) and a considerable decrease of elastic modulus (E) of the enamel. The prismatic structure of enamel was clearly observed after a one-hour immersion in Coca-Cola®, which shows its strong erosion effect. A high surface roughness of enamel results in a high chance of cavities due to easier bacterial adhesion on rougher surface, while a drastic deterioration of the mechanical properties of the enamel weakens its protection property. Our findings show the variation of enamel surface at the very beginning stage of etching process by acidic drinks, which can also be applicable to the etching mechanism of enamel surface by other sources.

## 1. Introduction

Tooth enamel is a masterpiece of biological mineralized tissues and has attracted great interest from material scientists and biologists[1]. Tooth enamel is a rigid, inert and acellular tissue covering the tooth crown[1]. As the most highly mineralized and hardest tissue in the human body, it consists of 96 wt% inorganic minerals, which are mainly well-organized carbonated hydroxyapatite crystals, and 2 wt% organic substances such as proteins and 2 wt% water[1]. Enamel rods (prisms) and inter-rods (interprismatic substance) represent the primary and fundamental structural units of tooth enamel. The enamel rods are approximately 5 μm diameter cylindrical-like structures[1,2]. The mechanical properties of enamel, e.g. the average elastic modulus of top surface of enamel, is reported in the range of 70 – 120 GPa[1,3].

As is well known, dental erosion can be induced by acidic foods, such as citrus fruits and acidic beverages due to demineralization[4-9]. Inorganic minerals and organic remnants can be dissolved by acidic solutions, which would result in roughened and weakened enamel surface. For example, Machado *et al.* reported the increased surface roughness and decrease of surface hardness and elastic modulus of enamel after beverage contact[7]. Even worse the roughened surface would cause more bacterial adhesion, colonization and then lead to cavities[8,9]. For example, Bollen *et al.* reported that surface roughness can influence bacterial plague retention[8]. Nogueira *et al.* also reported the increase of bacterial adhesion on tooth enamel due to the increase in surface roughness[9].

The early stages of enamel demineralization by acidic beverages are of high scientific relevance. Previous important studies in this field are very helpful to understand the deterioration phenomenon of enamel by acidic soft drinks. For example, Machado *et al.* detected a microscopic scale erosion on enamel surface after a 30-minute immersion time in acidic beverages by profilometer and nanoindentation method[9]. Kato *et al.* also reported a microscopic scale erosion phenomenon on enamel surface after 5 minutes of immersion in Coca-Cola® by scanning electron microscopy[10]. While these researches mainly focus on the roughening and softening of enamel surface in the microscale, nanoscale characterization of this phenomenon is still lacking. We know that every etching process begins in the nanoscale. Nanoscale characterization of these properties can help us to understand the very early stage of etching process and to elucidate the etching process clearer.

Atomic force microscopy (AFM) is one of the most suitable techniques to characterize the variation of the surface topography and mechanical properties quantitatively at the nanoscale[11-13]. The small radius of the cantilever tip (usually between 2 and 25 nm) allows for nanoscale characterization of both the morphology and mechanical properties of the tooth enamel. Different from literature that used AFM to characterize surface roughness ($R_a$) and hardness[14-18], here we use AFM to measure the surface roughness and elastic modulus (E) of teeth enamel immersed in various soft drinks over time and analysed the surface topography and elastic modulus map. The reason that we choose to characterize elastic modulus over hardness is that elastic modulus shows the resistance to elastic deformation while hardness is a measurement of the resistance to localized plastic deformation. An elastic response is non-permanent, while in the measurement of hardness, a plastic response leads to permanent deformation. As we intend to study the variation of mechanical property in very early stage of the enamel softening when they are immersed in acid environment, we think that elastic modulus is an excellent marker to trace the very beginning stage of the weakening of tooth enamel during the etching process. In addition, elastic modulus is linked to the density of the materials, so it will provide insight into the distribution of materials density as etching proceeds. Finally, we discussed the potential impact on the community of this research.

## 2. Materials and Methods

### 2.1 Preparation of the specimens

This study was approved by the Institutional Review Board (IRB: KH2018-82) of the Korea Advanced Institute of Science and Technology (KAIST). All research was performed in accordance with relevant guidelines and regulations. Furthermore, informed consent was obtained from all participants. Five healthy human molar teeth were obtained from volunteers between age 20 and 35 who visited the KAIST clinic. After extraction from volunteers, the teeth were preserved in distilled water before the experiment. To prepare the enamel samples for the experiments, the teeth were cross-sectioned along the dash lines as shown in Figure 1 (a). Teeth (molar) were cut into 2 mm thick slices to fit the AFM liquid cell. A top view of a slice is shown in Figure 1(b). Before the experiment, the specimens were polished by sand paper (grit #2000) and abrasive cloth. Then, the slice was adhered to the AFM liquid cell by Loctite adhesive leaving the polished side exposed to the liquid shown in Figure 1(c). In the present study, we confined the region of interest to the enamel part of the teeth from the optical images from AFM to investigate its topography and elastic modulus distribution.

**2.2 Demineralization experiment**

Commercially available soft drinks, including Coca-Cola®, Sprite® and orange juice (Minute Maid®), were purchased and opened right before the immersion experiment. NanoWorld Pointprobe® DH-NCHR silicon probes with an average spring constant of 80 N/m and a resonance frequency of approximately 400 kHz were used. Fluid Cell Lite (Oxford Instruments, Asylum Research) was used for carrying out the demineralization experiment of enamel.

To observe the morphology and mechanical property variation with etching time in soft drinks by AFM, it is necessary to discontinue etching at a certain time and make a series of specimens to see the whole chronological etching process. The surface morphology was measured by AC mode (tapping mode) AFM at a scan rate of 1 Hz. The F-map image was measured by contact force mode AFM. 2 ml of soft drink were transferred into the AFM liquid cell by pipette to fully cover the specimen. After 1 minute, the soft drink was taken out by pipette, and the specimen was kept inside the cell and gently rinsed with distilled water. The topography of the surface after a 1-minute etching was characterized by tapping mode AFM using the same scan rate and scan size. The F-map image was taken by contact force mode. By following the same procedure, the etching was repeated, and the surface topography was taken at each time point (3, 5 and 10 minutes). The topography of the enamel surface etched by Coca-Cola® for one hour was characterized by AC mode (tapping mode) AFM as well.

## 2.3 Atomic force microscopy imaging and spectroscopy

A commercially available atomic force microscopy (MFP-3D Origin, Asylum Research, Oxford Instruments) was used to characterize the surface roughness and elastic modulus of the specimens after various immersion times. NanoWorld Pointprobe® DH-NCHR silicon probes with an average elastic modulus (E) of 150 GPa and average spring constant of 80 N/m were used. The surface topography was obtained by AC mode (tapping mode) imaging, which minimizes wear and tear of the teeth surface. Surface roughness was calculated by the Asylum Research software based on the surface height images. The cantilever tip has a pyramid shape coated with diamonds with an elastic modulus of 500 – 1000 GPa. The maximum loading force of the load-displacement F-d curves is 7 μN. The indentation depth varied from several to tens of nanometers depending on the different softening stages of the enamel surface. The cantilever probe is supposed to detect the enamel layer in Figure 1(b). The elastic modulus map shown in Figure 4 was calculated by the Hertzian Model through data analysis of the load-displacement F-d curves on F-map by Asylum Research software based on the Equations 1 and 2. Although the Hertzian Model is usually used for calculating spherical probes, the assumption here is a spherical contact due to the very small indentation depth.

## 3. Results and Discussion

In the present study, we used three different soft drinks, Coca-Cola®, Sprite® and orange juice (Minute Maid®), to monitor the etching process of human teeth using AFM liquid cell. Figure 1 shows how we cut the teeth and attached the teeth slice to AFM liquid cell. The pH values of the soft drinks were measured just before the experiment, for which the values and compositions are summarized in Table 1. Coca-Cola® has the lowest pH, while orange juice has the highest pH. Previous studies have already shown that acids, such as citric acid and phosphoric acid, are the main cause of deterioration of the mechanical properties of teeth enamel[5,6,9]. After analyzing the ingredients of these soft drinks, we presume that phosphoric acid is the main etching cause of Coca-Cola®, while citric acid is the main etching cause of Sprite® and orange juice (Minute Maid®).

Figure 2(a) shows the variation of the enamel surface topography in the etching process of different soft drinks characterized by AFM. We observed that the surface topography underwent significant changes as the etching time increased. First, as the etching time increased, we started to see the emergence of a spherical particulate structure as well as some indication of a prismatic structure from the topography of the teeth enamel. Second, we found preferential etching of the polishing lines. Compared with the as-polished sample, as the etching time increased, the polishing lines became much wider. This indicates that if we brush our teeth too hard or use some toothpaste that has polishing particles to make some scratches, the scratched lines of teeth enamel can be etched away faster.

For comparison, the variation of the surface roughness is plotted in Figure 2(b). As can be observed clearly, the biggest change in $R_a$ was observed in the teeth exposed to Coca-Cola®. Orange juice had a relatively slower etching rate compared to Coca-Cola® and Sprite®. The $R_a$ of the enamel surface increased from 17 nm to 75 nm with the increase in etching time from 0 to 10 minutes for Coca-Cola®. What is more, $R_a$ increased more than 50% in the first minute, suggesting a fast surface roughing at the initial stage. As mentioned before, a higher surface roughness causes easier bacteria adhesion and colonization, which lead to cavities[8,9]. From this experiment, we found that soft drinks can cause an obvious enamel surface change in the nanoscale even in a very short time. Our results are in good agreement with similar research measured by AFM. For example, Watari *et al.* reported obvious surface roughness increase of polished enamel surface after 1 miniute immersion in 2% phosphoric acid or 10% citric acid by AFM[12].

To determine how serious it is when we soak our teeth in soft drinks, teeth slices were put into Coca-Cola® for a 1 h immersion. The topography of the enamel surface etched by Coca-Cola® for one hour was characterized by AFM AC-mode with a scan rate of 1 Hz as shown in Figure 3. The prismatic structure, which is also called "enamel rod," has been revealed on enamel surface after a one-hour demineralization in Coca-Cola®. Figure 3 (c) shows the height profile along the red line in Figure 3 (b). The enamel rod is measured to have a diameter of 5 μm, which is in good agreement with prior reports[19, 20]. This shows that the substance in the inter-rod (which are mainly organic phase including proteins) is much easier to be removed by Coke than enamel rod itself (which are mainly composed of hydroxyapatite crystals).

From the above observations, we can conclude that at the nanoscale level, the etching process of teeth begins very early when exposed to the soft drinks and continues over time. When the etching period is over one hour, the enamel prismatic structure will be revealed. Once these prismatic structures are exposed, bacteria and acids will go through the prismatic structures and reach the dentin, which will cause our teeth to become more sensitive to heat, cold, sweet and sour foods and even become hypersensitive[21, 22]. Therefore, if the immersion period in soft drinks is too long, serious and irreversible damage to the enamel will happen.

It is very important to know how soft drinks influence the mechanical properties of enamel because enamel has the main responsibility of protecting the teeth; thus, we measured the variation in the elastic modulus of the enamel influenced by soft drinks, as shown in Figure 4.

Figure 4 (a) shows the elastic modulus (E) map of the enamel surface after immersing the human teeth in different soft drinks. All elastic modulus values were calculated by the Asylum Research software using force-distance curve measurements. First, a matrix of 8-by-8 force–distance curves was measured on an area of 1 μm by 1 μm. We chose the loading curves to calculate the elastic modulus as the unloading curves also contain information of adhesion properties. The elastic modulus was calculated from each indentation curve based on the Hertzian Model, as illustrated by Equation (1) below[11]:

$$\frac{F}{d^{3/2}} = \frac{4}{3} E \times R^{1/2} \qquad (1)$$

where F is the loading force, d is the indentation depth, R is the tip radius. The E calculated from the above equation is actually $E_r$, the reduced modulus. The sample elastic modulus $E_s$ is calculated by Equation (2) below:

$$\frac{1}{E_r} = \frac{1 - v_i^2}{E_i} + \frac{1 - v_s^2}{E_s} \qquad (2)$$

where $E_i$ is the modulus of indenter. $v_s$ and $v_i$ are the Poisson's ratio of sample and indenter, respectively. $E_s$ is the sample elastic modulus we want to know. The obtained elastic modulus values were displayed as a coloured pixel on the elastic modulus map.

As shown in the scale bar, the measured E values are in the range of 0 – 150 GPa. A darker colour indicates a lower E while a brighter colour means a higher E. The E map revealed an inhomogeneous distribution of E on the enamel surface with a mixture of colours. The overall colour becomes darker and darker when increasing the immersion time, suggesting that immersion in soft drinks decreased the elastic modulus of the whole enamel surface.

Four typical force-indentation curves of the teeth that were immersed in Coca-Cola® are shown in Figure 4 (b). We can see that while the maximum loading force of all these curves on the enamel side of the teeth are controlled at 7 μN, the indentation depths are 20 and 70 nm for the as-polished enamel and the one immersed in Coca-Cola® for 5 minutes, respectively. This clearly shows an increase in the indentation depth for the enamel immersed in Coca-Cola®. The elastic modulus determined by AFM nanoindentation of as-polished enamel agrees with magnitudes derived previously from other techniques, such as instrumented nanoindentation[9]. Moreover, we found that the slope of the indentation curve (loading curve) decreased as a function of the etching time. The average E as a function of the etching time for different beverages is plotted in Figure 4 (c). The mean value of E for the enamel surface decreased from *ca*. 100 GPa to less than 10 GPa after 5 minutes of immersion in soft drinks. Furthermore, the elastic modulus decreased by around 20% in the first minute from an average of 100 GPa to 80 GPa, indicating a fast surface softening in the first minute. The elastic modulus dropped to less than 10% of its original value after a 5-minute immersion in the soft drinks. This agrees well with the trend for the $R_a$ measurement. Both results confirm that Coca-Cola® is the most effective soft drink in changing the surface roughness and elastic modulus of teeth.

Many researchers have already confirmed the negative effect of carbonated drinks on human enamel[14-18], but different from the previous reports, *our research shows that in the very early stage of the etching process, increase of the surface roughness and deterioration of elastic modulus at the nanoscale level can be observed*. Once the enamel is weakened, the chewing and biting activity will easily remove the weakened enamel surface. Our findings are significantly important in the following aspects. 1) The influence of soft drinks to enamel can be observed earlier in the nanoscale than previous microscopic scale studies. 2) As the main source of etching comes from acids, in some dental treatment procedures using acidic solutions (in the case of dental scaling or acid etching technique related to restorative dentistry), it is suggested that acid should be well controlled and minimized in dental treatment to avoid severe erosion. 3) Toothpastes with polishing particles are advertised to remove dental biofilms; however, it will cause scratches on the enamel surface, which can be preferential sites for etching. 4) Although fruits are healthy because they contain a lot of Vitamin C, acidic fruits such as oranges and lemons containing citric acid are considered most damaging to teeth enamel.

## 4. Conclusions

By using atomic force microscopy, we quantitatively evaluated the variation of the surface roughness ($R_a$) and elastic modulus (E) of human teeth enamel caused by commercially available beverages (Coca-Cola®, Sprite® and orange juice (Minute Maid®)) over different periods at the nanoscale level. We found that the surface of human teeth gradually roughened while in contact with soft drinks, and the surface roughness ($R_a$) increased linearly with the etching time. Furthermore, the elastic modulus (E) dropped tremendously with different soaking times in the soft drinks. It is worth noting that the roughening of the enamel surface and the degradation of the enamel strength in terms of elastic constant occur in a short time with beverage contact. Our findings suggest a higher chance of cavities due to the nanoscale deterioration of the enamel layer when exposed to various soft drinks.

## Acknowledgements

Dr. Panpan Li was supported by the Korea Research Fellowship Program funded by the National Research Foundation of Korea (no. 2017H1D3A1A01054478). The work was also supported by the KAIST GCORE (Global Center for Open Research with Enterprise) grant funded by the Ministry of Science and ICT (Project #N11180136) and KUSTAR-KAIST Institute, KAIST, Korea. We gratefully acknowledge Dentist Dr. Suebean Cho in the KAIST clinic for collaborating on this research. The authors greatly acknowledge Dr. Sangmin Shin at Smile Well Dental for his advice and stimulating discussion. We also gratefully acknowledge the suggestions and advice from Professor Kack-Kyun Kim in the School of Dentistry at Seoul National University.

**Author contributions**

P.L. and S.H. conceived and designed the experiments. P.L. carried out the experiments. C.O. and H.K. helped with the experiments. G.P. and H.K. helped with the discussion. P.L. and S.H. wrote the manuscript. All the authors discussed the results and commented on the manuscript.

**Declaration of interest**

Declarations of interest: none.

**Figure Legends and Table**

**Figure 1**. Schematic diagram of the cutting method (a) to get the tooth slices (b) and image of a tooth slice mounted on the AFM liquid cell (c).

**Figure 2**. (a) Variation of the enamel surface topography with different immersion times treated with Coca-Cola®, Sprite® and Orange juice (Minute Maid®). Scale bars, 2 μm. (b) Change of surface roughness with different immersion times in soft drinks.

**Figure 3**. (a) (b) Topography of the enamel prismatic structure after a 1-hour immersion in Coca-Cola®. (c) Height profile along the red line in (b).

**Figure 4**. (a) Variation of the enamel's elastic modulus (E) with different immersion times in Coca-Cola®, Sprite® and Orange juice (Minute Maid®). (b) Typical indentation curves of the enamel samples for the as-polished and the different immersion times treated by Coca-Cola®. (c) Change of E with the different immersion times in the soft drinks.

**Table 1**. Ingredients and pH values of the soft drinks.

**Table 1. Ingredients and pH values of the soft drinks.**

|  | pH value at room temperature | Main components[23] |
|---|---|---|
| Coca-Cola® | 2.23 at 25.0 °C | Carbonated water, high fructose corn syrup, Caramel color, Phosphoric acid, Natural flavors, Caffeine. |
| Sprite® | 3.03 at 25.7 °C | Carbonated water, high fructose corn syrup, citric acid, natural flavors, sodium citrate, sodium benzoate. |
| Orange juice (Minute Maid®) | 3.49 at 20.5 °C | organic acids (citric, malic, and ascorbic acid), sugars, and phenolic compounds. |

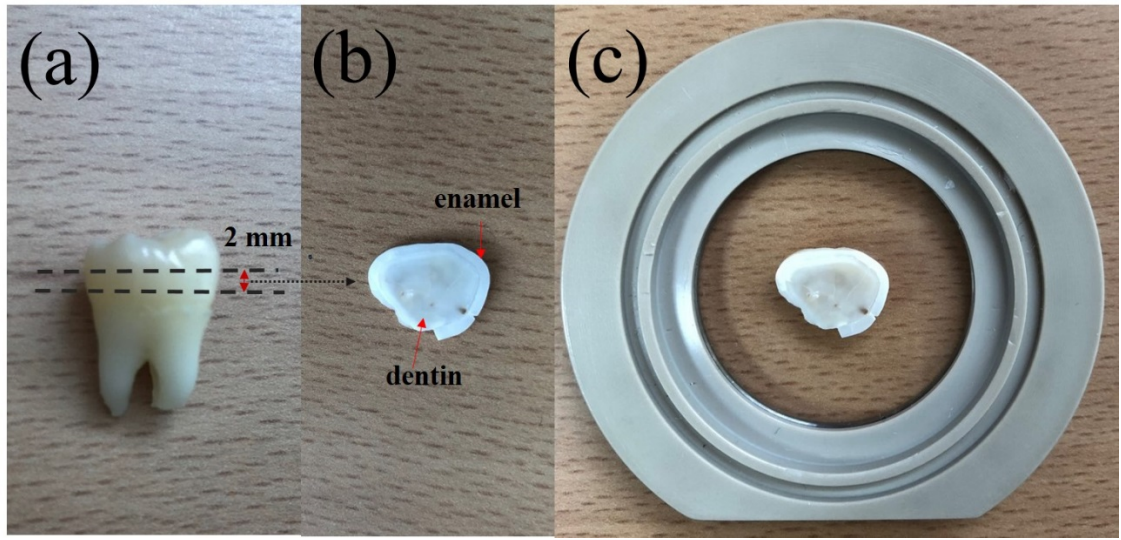

**Figure 1.** Photos taken by a camera, (a) shows we cut the tooth along the direction indicated by the dash lines, (b) shows the cross section of the 2 mm tooth slice, and (c) shows how tooth slice was mounted on the AFM liquid cell.

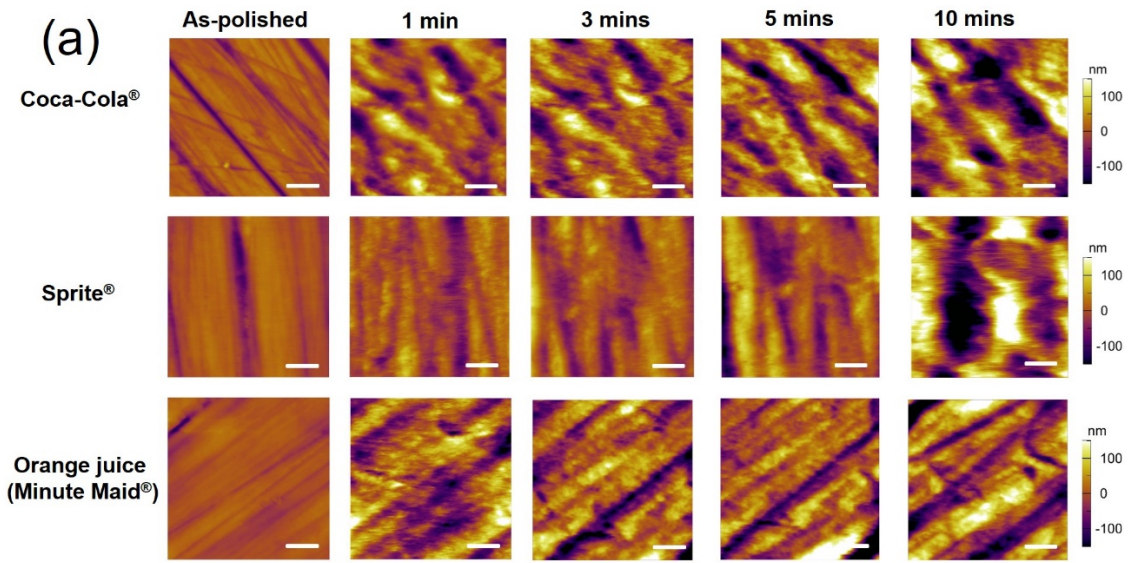

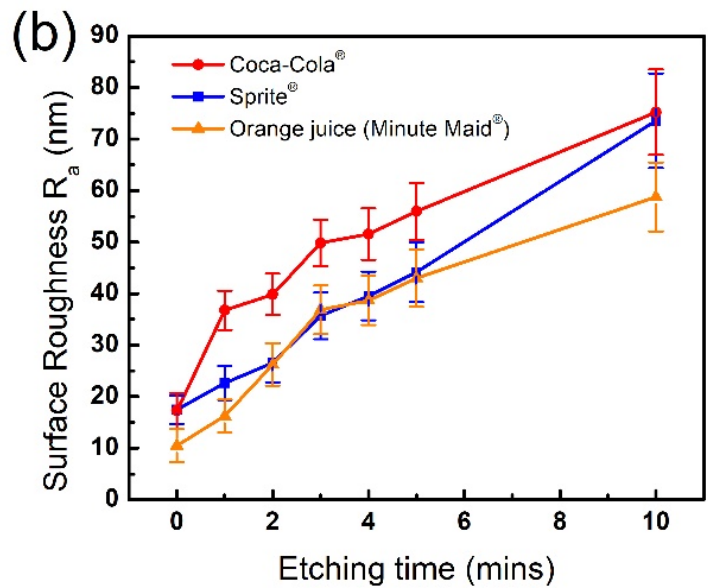

**Figure 2.** (a) Variation of the enamel surface topography with different immersion times treated with Coca-Cola®, Sprite® and Orange juice (Minute Maid®). Scale bars, 2 μm. (b) Change of surface roughness with different immersion times in soft drinks.

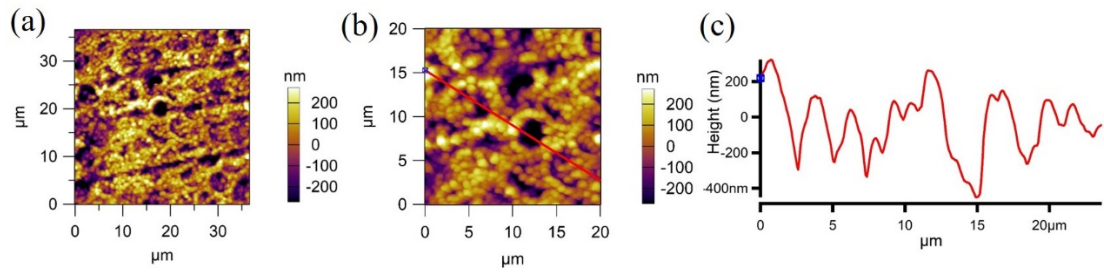

**Figure 3.** (a) (b) Topography of the enamel prismatic structure after a one-hour immersion in Coca-Cola®. (c) Height profile along the red line in (b).

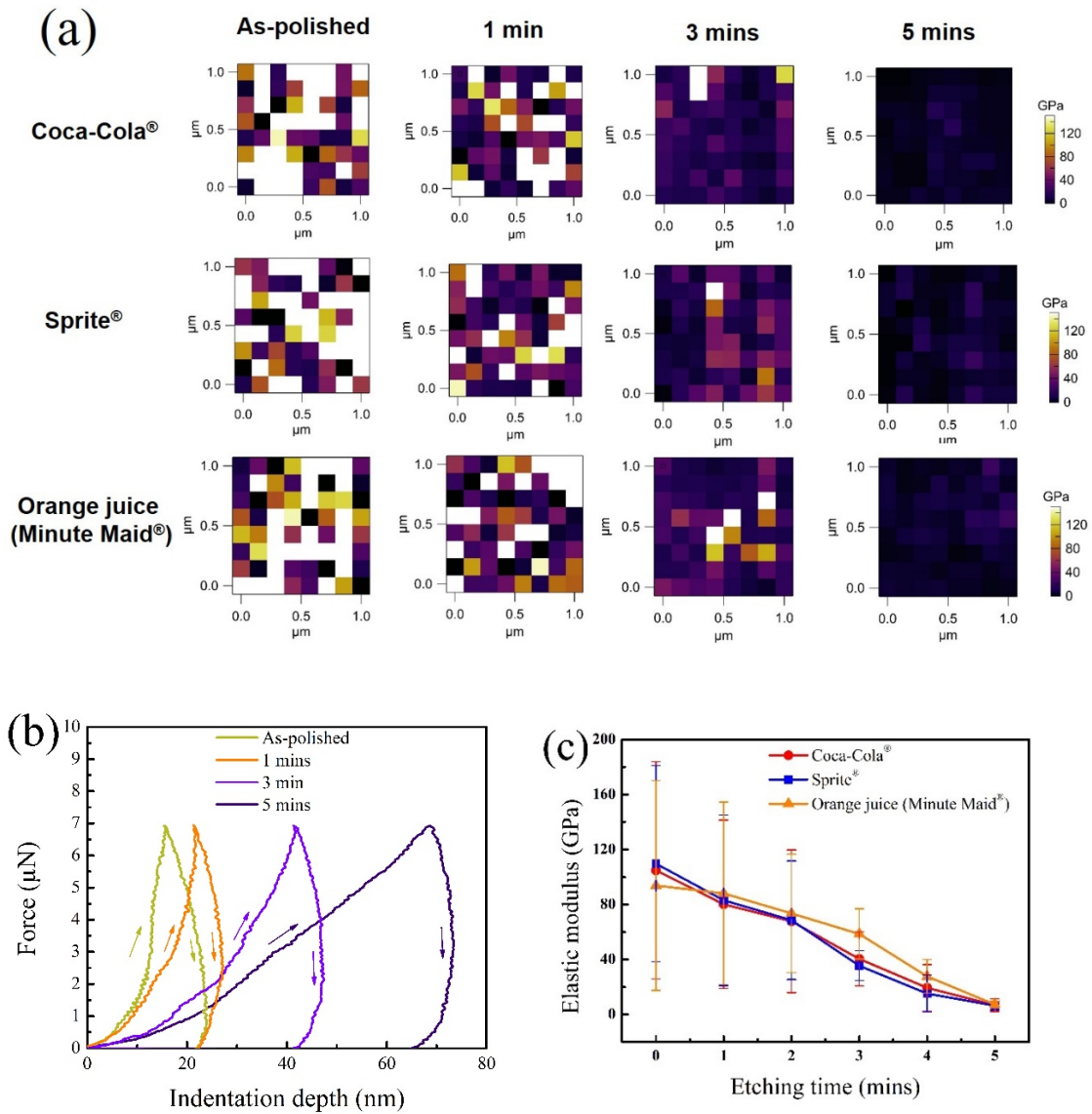

**Figure 4.** (a) Variation of the enamel's elastic modulus (E) with different immersion times in Coca-Cola®, Sprite® and Orange juice (Minute Maid®). (b) Typical indentation curves of the enamel samples for the as-polished and the different immersion times treated by Coca-Cola®. (c) Change of E with the different immersion times in the soft drinks.